\documentclass[conference,twocolumn,10pt]{IEEEtran} 
\usepackage[utf8]{inputenc}
\usepackage{epsfig,cite}
\usepackage{graphicx}
\graphicspath{{graphics/}}
\usepackage{color}
\usepackage{amssymb,amsmath}
\usepackage{mathrsfs} 

\IEEEoverridecommandlockouts

\DeclareMathOperator{\erf}{erf}
\DeclareMathOperator*{\argmax}{arg\,max}

\begin{document}
	
	\title{Asynchronous Peak Detection for \\Demodulation in Molecular Communication}
	
	\author{
		\IEEEauthorblockN{Adam Noel}
		\IEEEauthorblockA{School of EECS\\ University of Ottawa\\ Ottawa, Ontario, Canada\\ Email: anoel2@uottawa.ca}
		\and
		\IEEEauthorblockN{Andrew W. Eckford}
		\IEEEauthorblockA{Dept. of EECS\\
			York University\\
			Toronto, Ontario, Canada\\
			Email: aeckford@yorku.ca}
		
		\thanks{This work was supported in part by the Natural Sciences and Engineering Research Council of Canada (NSERC).}
		
	}

	\newcommand{\EXP}[1]{\exp\left(#1\right)}
	\newcommand{\ERF}[1]{\erf\left(#1\right)}
	
	\newcommand{\obsSignal}[1]{y[#1]}
	\newcommand{\obsSignalExp}[1]{\overline{y}[#1]}
	\newcommand{\obsSignalExpISI}[1]{\overline{y}_\textnormal{ISI}[#1]}
	\newcommand{\obsSignalOne}[1]{z_\ell[#1]}
	\newcommand{\obsSignalOneExp}[1]{\bar{z}_\ell[#1]}
	\newcommand{\obsInd}{k}
	\newcommand{\obsIndRX}{j}
	\newcommand{\obsIndTwo}{i}
	\newcommand{\obsOffset}{\delta}
	
	\newcommand{\pmf}[1]{f_{#1}}
	\newcommand{\probSingleObs}[1]{p[#1]}
	\newcommand{\prob}[1]{P_\textnormal{#1}}
	\newcommand{\rv}{X}
	\newcommand{\rvVal}{x}
	\newcommand{\rvAvg}{\overline{x}}
	\newcommand{\maxSample}[1]{m_{#1}}
	\newcommand{\maxSampleAdaptive}[1]{\tilde{m}_{#1}}
	\newcommand{\maxSampleVal}{a}
	
	\newcommand{\dataLength}{L}
	\newcommand{\dataInd}{l}
	\newcommand{\dataIndOther}{n}
	\newcommand{\bit}[1]{b_{#1}}
	\newcommand{\bitSeq}{\mathbf{b}}
	\newcommand{\bitObs}[1]{\hat{b}_{#1}}
	\newcommand{\bitSeqObs}{\hat{\mathbf{b}}}
	\newcommand{\obsPerBit}{M}
	\newcommand{\threshold}{\tau}
	\newcommand{\pOne}{P_1}
	\newcommand{\pZero}{P_0}
	\newcommand{\pError}{P_\textnormal{e}}
	\newcommand{\pErrorCur}[1]{\pError[#1]}
	\newcommand{\pErrorAvg}{\overline{P}_\textnormal{e}}
	\newcommand{\sumAdaptive}[1]{\tilde{\alpha}_{#1}}
	
	\newcommand{\molReleased}{N}
	\newcommand{\diffusion}{D}
	\newcommand{\dt}{\Delta t}
	\newcommand{\rrx}{r_\textnormal{RX}}
	\newcommand{\Vrx}{V_\textnormal{RX}}
	\newcommand{\dist}{d}
	\newcommand{\metre}{\textnormal{m}}
	\newcommand{\second}{\textnormal{s}}
	\newcommand{\mol}{\textnormal{mol}}
	
	\maketitle

	\begin{abstract}
		Molecular communication requires low-complexity symbol detection algorithms to deal with the many sources of uncertainty that are inherent in these channels. This paper proposes two variants of a high-performance asynchronous peak detection algorithm for a receiver that makes independent observations. The first variant has low complexity and measures the largest observation within a sampling interval. The second variant adds decision feedback to mitigate inter-symbol interference. Although the algorithm does not require synchronization between the transmitter and receiver, results demonstrate that the bit error performance of symbol-by-symbol detection using the first variant is better than using a single sample whose sampling time is chosen \emph{a priori}. The second variant is shown to have performance comparable to that of an energy detector. Both variants of the algorithm demonstrate better resilience to timing offsets than that of existing detectors.
	\end{abstract}

	\maketitle
	
	\section{Introduction}
	
	In molecular communication, signals are conveyed from transmitter to receiver in patterns of molecules, which propagate via Brownian motion; see \cite{Nakano2013c}. For example, information can be conveyed in the quantity of molecules. A single bit $\{0,1\}$ may be transmitted by releasing zero molecules for $0$, or a large number of molecules for $1$. In this example, the receiver's task is to observe the number of arriving molecules and determine which bit was sent.
	
	Presently, molecular communication design follows the paradigm of conventional communication systems; signal detection techniques often assume synchronization and channel state knowledge. Algorithms to accomplish these tasks are an active area of research. For example, work has been done to address synchronization (see \cite{Abadal2011a,Moore2013,Shahmohammadian2013,Lin2016}) and parameter estimation (see \cite{Moore2012,Noel2014c}) in molecular communication. 
	Synchronization in particular is an important issue; in the opening example above, the transmitter and receiver must agree on when to transmit and when to detect. Many contemporary papers assume that perfect synchronization can be achieved, particularly when timing is used to convey information; see \cite{Rose2015,Srinivas2012}. On the other hand, asynchronous symbol detection is also an active research area in molecular communication. For example, in \cite{Lee2015a}, the propagation time is estimated from the variance of arrival times of a large number of molecules. In \cite{Lin2015b}, molecules arrive over time and a counter increments until a threshold number is observed.
	
\begin{figure}[!tb]
		\centering
		\includegraphics[width=\linewidth]{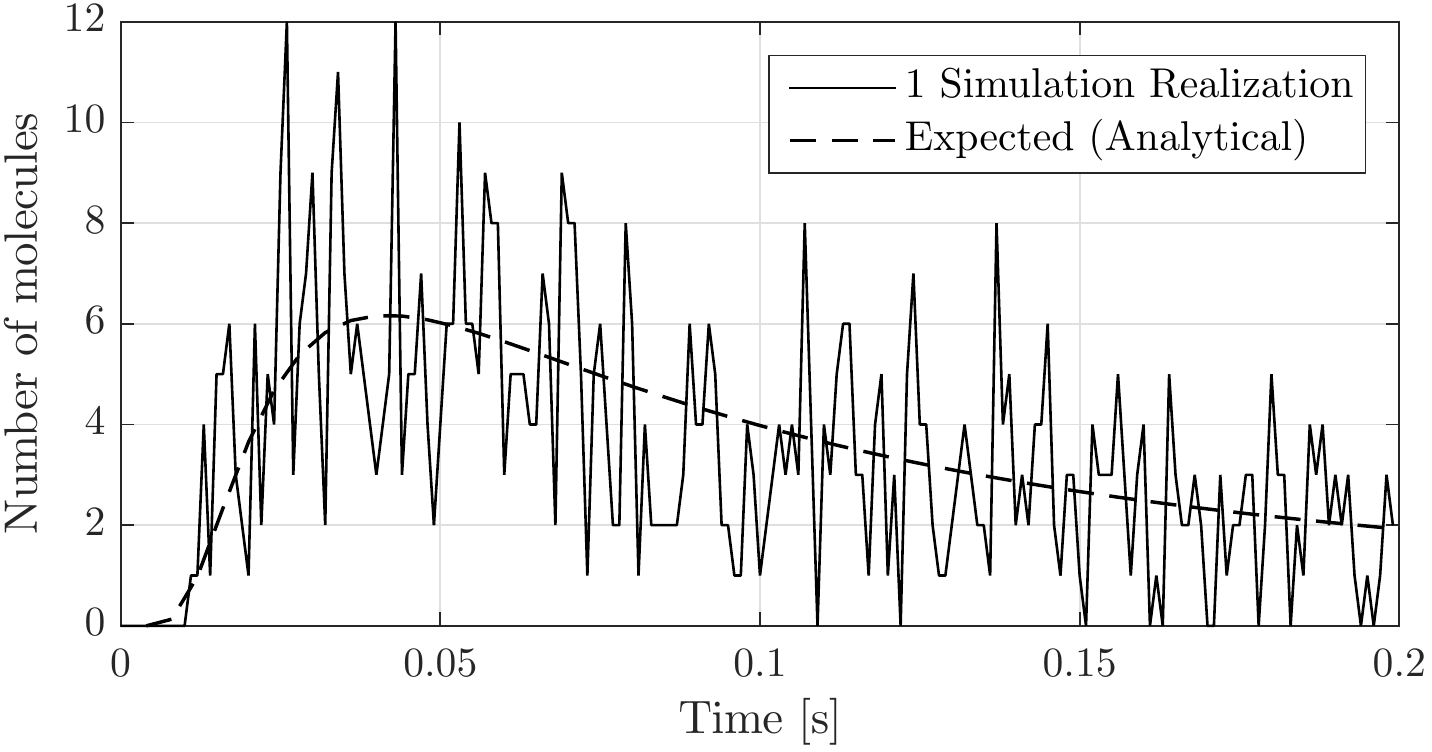}
		\caption{Channel impulse response for a passive receiver. The environment considered here is the same as that in Section~\ref{sec_results}, except the sampling period is decreased to $1\,\metre\second$. One sample realization of the received signal is compared with that expected from the analytical expression in (\ref{eqn_cir}), which has been scaled by the number of molecules $\molReleased$ released.}
		\label{fig_cir}
	\end{figure}

	Complex synchronization algorithms are not practical for nanonetworking applications, which envision low-complexity nanoscale devices (or biological ``devices'') whose computational capabilities are limited; see \cite{Akyildiz2008}. This paper focuses on \emph{asynchronous peak detection} for demodulation, which we also supplement with decision feedback as a variant. To motivate our design, consider the scenario in Fig.~\ref{fig_cir}. At time $t = 0$, a number $\molReleased$ of molecules is released by the transmitter, representing a bit $1$. The transmitter and receiver agree on the communication strategy, but are not synchronized; the receiver distinguishes between a $0$ and a $1$ by observing the {\em peak} of the response. If $\molReleased \rightarrow \infty$, we would observe something close to the dashed line in the figure, i.e., a relatively smooth response with a clear peak. However, in nanonetworking applications, $\molReleased$ is relatively small. Thus, the figure shows a simulated curve with many local peaks, both before and after the peak of the expected curve. Our goal is to detect the peak asynchronously and in the presence of inter-symbol interference (ISI).

	Related work in this direction includes \cite{Li2016a,Damrath2016}, which presented non-coherent detection algorithms that require no knowledge of the underlying channel impulse response. In \cite{Li2016a}, the local convexity of the diffusive signal is exploited. In \cite{Damrath2016}, an adaptive threshold detector subtracts the previous observation from the current sample. In this work, we simply find the largest sample, which results in a simple yet effective asynchronous detector. However, we do apply the expected impulse response (asynchronously) in the variant with decision feedback.
	
	The main contribution of this paper is the design and analysis of the asynchronous peak detection scheme for diffusive signaling, where the symbol is modulated by the quantity of molecules and detected by observing the size of the peak. Unlike other approaches to this problem, we explicitly model the timing offset between the transmitter and receiver, and the variant with (asynchronous) decision feedback mitigates the ISI that is common in molecular communication systems. Our proposed design significantly outperforms the common single-sample detector, and, in the presence of a large timing offset, the variant with decision feedback can outperform an energy detector with decision feedback.
	
	The rest of this paper is organized as follows. In Section~\ref{sec_model}, we present our asynchronous system model and the distribution of the largest observation in a given sampling interval. In Section~\ref{sec_rx}, we propose and analyze the asynchronous peak detector. We verify the analysis by comparing expected detector performance with simulations in Section~\ref{sec_results}, and conclude in Section~\ref{sec_concl}.
	
	\section{System Model and Preliminaries}
	\label{sec_model}
	
	\subsection{System Model}
	
	We consider a diffusion-based molecular communication system with one transmitter and one receiver. The transmitter releases molecules according to a random binary sequence $\bitSeq = [\bit{0}, \bit{1},\ldots,\bit{\dataLength-1}]$. Time is divided into slots of size $\dt$ and is referenced by index $\obsInd \in \{0,1,\ldots\}$. For simplicity, the transmitter modulates with impulsive ON/OFF keying and begins transmitting at time index $\obsInd = 0$. Each bit lasts $\obsPerBit$ slots: every $\obsPerBit$ slots, the transmitter releases $\molReleased$ molecules for a bit 1 and no molecules for a bit 0. 
	The receiver makes discrete observations $\obsSignal{\obsInd-\obsOffset} = \obsSignal{\obsIndRX}$, where $\obsOffset$ is some constant but unknown offset from the transmitter's clock. From the observations $\obsSignal{\obsIndRX}$, the receiver detects the sequence $\bitSeqObs = [\bitObs{0}, \bitObs{1},\ldots,\bitObs{\dataLength-1}]$. Although this results in blocks of length $\dataLength$, we do not consider blockwise detection in this paper, as we focus on low-complexity symbol-by-symbol algorithms.
	
	Generally, we make no assumptions about the geometry of the environment, including the shape of the transmitter and receiver, nor about the reception mechanism at the receiver. We do assume that the receiver's observations are \emph{independent}, such that $\obsSignal{\obsIndRX}$ is only conditioned on the expected time-varying signal $\obsSignalExp{\obsIndRX}$ and not on $\obsSignal{\obsIndTwo}$, where $\obsIndTwo\ne\obsIndRX$.  It is also possible to extend our analysis to a different modulation scheme or to release the $\molReleased$ molecules over a finite interval.
	
	Two examples of the received signal $\obsSignal{\obsIndRX}$ are as follows. If the receiver is a passive observer, then $\obsSignal{\obsIndRX}$ is the number of molecules at the receiver at time $\obsIndRX\dt$. If the receiver has an absorbing surface, then $\obsSignal{\obsIndRX}$ is the number of molecules that have been absorbed within the interval $\big((\obsIndRX-1)\dt,\obsIndRX\dt\big]$. Strictly speaking, neither of these signals have independent observations, but we leave a detailed consideration of sample dependence for future work.
	
	\subsection{Receiver Signal}
	
	When the transmitter releases $\molReleased$ molecules, we assume that the diffusive behavior of each individual molecule is independent. Using the independence of molecule behavior and receiver observations, and given that molecules are only released at transmitter time indices $\obsInd$ that are integer multiples of $\obsPerBit$, then the observed number of molecules $\obsSignal{\obsInd}$ is a discrete-time random process that follows a Binomial distribution with $\molReleased$ trials and success probability $\probSingleObs{\obsInd}$.

	Let $z_\ell[k]$ represent the number of molecules observed at time $\obsInd$ due to a \emph{single} release of $\molReleased$ molecules at $\obsInd=\ell$, and assuming that no molecules were released at any other time. We note that, like $\obsSignal{\obsInd}$, $z_\ell[k]$ is a discrete-time random process.
	
	When the transmitter is modulating the binary sequence $\bitSeq$, then an observation at arbitrary time $\obsInd$ can include molecules from any current or previous symbol, i.e.,
	\begin{equation}
		\obsSignal{\obsInd} = \sum_{\dataInd=0}^{\lfloor\obsInd/\obsPerBit\rfloor} \bit{\dataInd}z_\ell[k],
		\label{eqn_rx_obs_sync}
	\end{equation}
	where the $z_\ell[k]$ term is associated with $\molReleased$ trials and success probability $\probSingleObs{\obsInd-\dataInd\obsPerBit}$ when $\bit{\dataInd} = 1$, and $\lfloor\cdot\rfloor$ is the floor function. Thus, $\obsSignal{\obsInd}$ is a summation of Binomial random variables with \emph{different} success probabilities, i.e., a Poisson Binomial random variable. Instead of evaluating the Poisson Binomial distribution (which has combinatorial complexity; see \cite{Hong2013}), it is computationally simpler to approximate each Binomial random variable as a Poisson random variable with mean $E[z_\ell[k]] = \bar{z}_\ell[k] =\molReleased\probSingleObs{\obsInd-\dataInd\obsPerBit}$. Thus, by the properties of Poisson random variables, $\obsSignal{\obsInd}$ is also a Poisson random variable with mean
	\begin{equation}
		\obsSignalExp{\obsInd} = \sum_{\dataInd=0}^{\lfloor\obsInd/\obsPerBit\rfloor} \bit{\dataInd}\bar{z}_\ell[k],
		\label{eqn_rx_obs_sync_poisson}
	\end{equation}
	which is accurate for $\molReleased$ sufficiently large and each $\probSingleObs{\obsInd-\dataInd\obsPerBit}$ sufficiently small; see \cite[Ch.~5]{Ross2009}. The cumulative distribution function (CDF) of Poisson random variable $\rv$ can be written as \cite[Ch.~24]{Abramowitz1964}
	\begin{equation}
		\Pr\{\rv \le \rvVal\} = \frac{\Gamma(\lfloor\rvVal+1\rfloor,\rvAvg)}{\Gamma(\lfloor\rvVal+1\rfloor)},
		\label{eqn_poisson_cdf}
	\end{equation}
	where $\Gamma(\cdot)$ and $\Gamma(\cdot,\cdot)$ are the Gamma and incomplete Gamma functions, respectively, and $\rvAvg$ is the mean of random variable $\rv$. Generally, for discrete $\rvVal$, we have $\Pr\{\rv \le \rvVal\} = \Pr\{\rv < \rvVal+1\}$. The form in (\ref{eqn_poisson_cdf}) also enables us to consider non-integer $\rvVal$, in which case $\Pr\{\rv \le \rvVal\} = \Pr\{\rv < \rvVal\}$. This property will be helpful when we consider the performance of detectors with decision feedback.
		
	Given the distribution of $\obsSignal{\obsInd}$, we can describe the distribution of the maximum observation over some interval. Specifically, for the $\dataInd$th bit, the receiver makes observations $\obsSignal{\obsIndRX}, \obsIndRX \in \{\dataInd\obsPerBit-\obsOffset,\ldots,(\dataInd+1)\obsPerBit-\obsOffset\}$. It can be shown that the maximum value of these $\obsPerBit$ observations, $\maxSample{\dataInd}$, has CDF
	\begin{equation}
		\Pr\{\maxSample{\dataInd} \le \maxSampleVal\} = \prod_{\obsIndRX = \dataInd\obsPerBit-\obsOffset}^{(\dataInd+1)\obsPerBit-\obsOffset} \Pr\{\obsSignal{\obsIndRX} \le \maxSampleVal\}.
		\label{eqn_max_cdf}
	\end{equation}
	
	In \cite{Noel2014c}, we used the maximum observation $\maxSample{\dataInd}$ to estimate channel parameters, but we did not describe its CDF nor use it for symbol detection, as we do in this paper.
	
	\subsection{Existing Detectors}
	
	To assess the performance of the asynchronous detector, we compare it with two existing detectors that are both variants of the weighted sum detectors that we proposed in \cite{Noel2014d}. The \emph{single sample detector} makes one observation at the time when the largest observation is \emph{expected}, i.e., at $\argmax \obsSignalExp{\obsInd}$ when $\bit{\dataInd}=1$, and compares that observation with the threshold $\threshold$. The \emph{energy detector} takes \emph{all} observations in the receiver interval, adds them together, and compares the sum with the threshold $\threshold$. The performance of these detectors is discussed in \cite{Noel2014d}, and we discuss the bit error probability of the energy detector with decision feedback in Section~\ref{sec_rx_ed}.
	
	\section{Receiver Design and Performance}
	\label{sec_rx}
	
	In this section, we propose the asynchronous peak detector. We consider two variants. The first variant (\emph{simple asynchronous detector}) is non-adaptive and compares each observation with the same constant threshold for every bit. The second variant (\emph{asynchronous detector with decision feedback}) is an adaptive detector that adjusts the threshold for \emph{each} observation in a bit interval, based on the inter-symbol interference (ISI) \emph{expected} in each observation. We derive the bit error probability of each variant, and also discuss the bit error probability of the energy detector with decision feedback.
	
	\subsection{Simple Asynchronous Detector}
	
	The simple asynchronous detector decodes the $\dataInd$th bit by comparing the maximum observation $\maxSample{\dataInd}$ with the constant threshold $\threshold$. Thus, the decision rule is
	\begin{equation}
		\bitObs{\dataInd} = \left\{
		\begin{array}{rl}
			1 & \text{if} \quad
			\maxSample{\dataInd} \ge \threshold,\\
			0 & \text{otherwise}.
		\end{array} \right.
	\end{equation}
	
	This detector is simple to implement because it only requires $\obsPerBit$ comparisons (including one with the threshold). We claim that this simple detector has significantly improved performance over a single sample detector for a wide range of offset $\obsOffset$, since this detector does not impose precisely when the maximum value should be observed. However, when $\obsOffset$ is sufficiently large (either positive or negative), the simple asynchronous detector has a higher risk of observing molecules transmitted in a different bit interval, due to the sampling window of length $\obsPerBit$. We leave the study of sampling window length for future work.
	
	For a given transmitter sequence $\bitSeq$, the average probability of error for the $\dataInd$th bit can be calculated as
	\begin{align}
		\pErrorCur{\dataInd} = & \pOne\Pr\{\maxSample{\dataInd} < \threshold | \bit{\dataInd}=1,\bit{\dataIndOther},\dataIndOther\ne\dataInd\}\, + \nonumber \\
		&\pZero\Pr\{\maxSample{\dataInd} \ge \threshold | \bit{\dataInd}=0,\bit{\dataIndOther},\dataIndOther\ne\dataInd\}.
		\label{eqn_simple_async_error}
	\end{align}
	
	To evaluate (\ref{eqn_simple_async_error}), we need $\Pr\{\maxSample{\dataInd} < \threshold\}$, and since $\threshold$ is discrete we use $\Pr\{\maxSample{\dataInd} < \threshold\} = \Pr\{\maxSample{\dataInd} \le \threshold-1\}$. We also find $\Pr\{\maxSample{\dataInd} \ge \threshold\}$ via $1-\Pr\{\maxSample{\dataInd} < \threshold\}$. From (\ref{eqn_max_cdf}), we need the distributions of the individual observations $\obsSignal{\obsIndRX}$, which we find via (\ref{eqn_poisson_cdf}) as
	\begin{equation}
		\Pr\{\obsSignal{\obsIndRX} \le \threshold-1\} = \frac{\Gamma(\lfloor\threshold\rfloor,\obsSignalExp{\obsIndRX})}{\Gamma(\lfloor\threshold\rfloor)},
		\label{eqn_obs_cdf}
	\end{equation}
	where the mean signal $\obsSignalExp{\obsIndRX}$ is evaluated via (\ref{eqn_rx_obs_sync_poisson}). In general, we need to account for ``future'' bits, i.e., $\dataIndOther > \dataInd$, when the offset $\obsOffset$ is negative. To evaluate the overall average probability of error $\pErrorAvg$ for a given threshold $\threshold$, we average (\ref{eqn_simple_async_error}) over all $\dataLength$ bits and all realizations of transmitter sequence $\bitSeq$.
	
	\subsection{Asynchronous Detector with Decision Feedback}
	\label{sec_rx_async_df}
	
	The asynchronous detector with decision feedback is an adaptive detector. It decodes the $\dataInd$th bit by first subtracting the expected ISI from every observation, or analogously adding the expected ISI to the constant threshold $\threshold$. Due to the time-varying nature of the channel impulse response, the expected ISI is \emph{also time-varying}. The decision rule is
	\begin{equation}
		\bitObs{\dataInd} = \left\{
		\begin{array}{rl}
			1 & \text{if} \quad
			\maxSampleAdaptive{\dataInd} \ge \threshold,\\
			0 & \text{otherwise},
		\end{array} \right.
	\end{equation}
	where $\maxSampleAdaptive{\dataInd}$ is the maximum adaptive observation, found as
	\begin{equation}
		\maxSampleAdaptive{\dataInd} = \max_{\obsIndRX \in \{\dataInd\obsPerBit-\obsOffset,\ldots,(\dataInd+1)\obsPerBit-\obsOffset\}} \obsSignal{\obsIndRX} - \obsSignalExpISI{\obsIndRX},
	\end{equation}
	and 
	\begin{equation}
		\obsSignalExpISI{\obsIndRX} = \sum_{\dataIndOther=0}^{\dataInd-1} \bitObs{\dataIndOther}\obsSignalOneExp{\obsIndRX+\obsOffset}
		\label{eqn_max_isi}
	\end{equation}
	is the average ISI expected by the receiver in the $\obsIndRX$th observation, conditioned on the receiver's previously-decided bits $\bitObs{\dataIndOther}, \dataIndOther \in \{0,1,\dataInd-1\}$. We emphasize that the receiver believes that its offset with the transmitter is $\obsOffset=0$, such that $\obsIndRX+\obsOffset=\obsInd$ in (\ref{eqn_max_isi}). This detector is more complex to implement than the simple asynchronous detector, since the receiver must have knowledge of $\obsSignalOneExp{\cdot}$ and be able to subtract terms of $\obsSignalOneExp{\cdot}$ from individual observations according to previous bit decisions.
	
	The corresponding CDF of the $\dataInd$th maximum adaptive observation $\maxSampleAdaptive{\dataInd}$ is
	\begin{equation}
		\Pr\{\maxSampleAdaptive{\dataInd} \le \maxSampleVal\} = \prod_{\obsIndRX = \dataInd\obsPerBit-\obsOffset}^{(\dataInd+1)\obsPerBit-\obsOffset} \Pr\{\obsSignal{\obsIndRX} - \obsSignalExpISI{\obsIndRX} \le \maxSampleVal\}.
		\label{eqn_max_adaptive_cdf}
	\end{equation}
	
	The average bit error probability of the asynchronous detector with decision feedback can be calculated using (\ref{eqn_simple_async_error}), where $\maxSample{\dataInd}$ is replaced with $\maxSampleAdaptive{\dataInd}$. Thus, we need the distribution of $\maxSampleAdaptive{\dataInd}$. Since $\obsSignalExpISI{\cdot}$ is generally non-integer, we should consider non-integer values of $\maxSampleVal$. However, the actual observations $\obsSignal{\cdot}$ are always discrete. So, we let $\maxSampleVal + \obsSignalExpISI{\obsIndRX} = \maxSampleVal[\obsIndRX]$ and consider the distribution of $\Pr\{\obsSignal{\obsIndRX} \le \maxSampleVal[\obsIndRX]\}$.
	
	We distinguish between the conditioning of $\obsSignal{\obsIndRX}$ and $\maxSampleVal[\obsIndRX]$. The observation $\obsSignal{\obsIndRX}$, whose mean $\obsSignalExp{\obsIndRX}$ is found via (\ref{eqn_rx_obs_sync_poisson}) using $\obsIndRX=\obsInd-\obsOffset$, is conditioned on the \emph{true} transmitter sequence $\bitSeq$ as well as the receiver's offset $\obsOffset$ from the transmitter's clock. The value of $\maxSampleVal[\obsIndRX]$, found using (\ref{eqn_max_isi}), is conditioned on the receiver's previous decisions $\bitObs{\dataIndOther}, \dataIndOther < \dataInd$, and the assumption that $\obsOffset=0$. From (\ref{eqn_poisson_cdf}), the CDF of $\obsSignal{\obsIndRX}$ is
	\begin{equation}
		\Pr\{\obsSignal{\obsIndRX} \le \maxSampleVal[\obsIndRX]\} = \frac{\Gamma(\lfloor\maxSampleVal[\obsIndRX]+1\rfloor,\obsSignalExp{\obsIndRX})}{\Gamma(\lfloor\maxSampleVal[\obsIndRX]+1\rfloor)},
		\label{eqn_obs_adaptive_cdf}
	\end{equation}
	and $\Pr\{\obsSignal{\obsIndRX} < \maxSampleVal[\obsIndRX]\} = \Pr\{\obsSignal{\obsIndRX} \le \maxSampleVal[\obsIndRX]\}$ when $\maxSampleVal[\obsIndRX]$ is non-integer. Thus, to use (\ref{eqn_max_adaptive_cdf}) to evaluate $\Pr\{\maxSampleAdaptive{\dataInd} < \threshold\}$, the $\obsIndRX$th observation CDF might be found using $\Pr\{\obsSignal{\obsIndRX} \le \threshold+\obsSignalExpISI{\obsIndRX}-1\}$ or $\Pr\{\obsSignal{\obsIndRX} \le \threshold+\obsSignalExpISI{\obsIndRX}\}$, depending on whether $\threshold+\obsSignalExpISI{\obsIndRX}$ is discrete or non-integer, respectively.
	
	\subsection{Energy Detector with Decision Feedback}
	\label{sec_rx_ed}
	
	We considered an energy detector with decision feedback in \cite{Noel2014e}, but we did not derive its expected bit error probability. The derivation is similar to that in Section~\ref{sec_rx_async_df}, where we observe the receiver's $\dataInd$th adaptive sum $\sumAdaptive{\dataInd}$ instead of its $\dataInd$th maximum adaptive observation $\maxSampleAdaptive{\dataInd}$, i.e.,
	\begin{equation}
		\sumAdaptive{\dataInd} = \sum_{\obsIndRX=\dataInd\obsPerBit-\obsOffset}^{(\dataInd+1)\obsPerBit-\obsOffset} (\obsSignal{\obsIndRX} - \obsSignalExpISI{\obsIndRX}),
	\end{equation}
	where $\obsSignalExpISI{\obsIndRX}$ has the same form as in (\ref{eqn_max_isi}), and the error probability is calculated using (\ref{eqn_simple_async_error}), where $\maxSample{\dataInd}$ is replaced with $\sumAdaptive{\dataInd}$. The CDF of $\sumAdaptive{\dataInd}$ has the same form as (\ref{eqn_obs_adaptive_cdf}), i.e.,
	\begin{equation}
		\Pr\{\sumAdaptive{\dataInd} \le \maxSampleVal\} = \frac{\Gamma(\lfloor\maxSampleVal[\dataInd]+1\rfloor,\sum_\obsIndRX\obsSignalExp{\obsIndRX})}{\Gamma(\lfloor\maxSampleVal[\dataInd]+1\rfloor)},
		\label{eqn_ed_adaptive_cdf}
	\end{equation}
	where $\maxSampleVal[\dataInd] = \maxSampleVal + \sum_\obsIndRX \obsSignalExpISI{\obsIndRX}$ and this term could be discrete or non-integer.
	
	\section{Numerical and Simulation Results}
	\label{sec_results}
	
	In this section, we evaluate the bit error performance of the proposed asynchronous detector in comparison with existing detectors. We consider both the simple variants and the adaptive variants with decision feedback. The analytical expressions are verified by comparing with particle-based simulations executed in the AcCoRD simulator; see \cite{Noel2016}.
	
	We consider an unbounded 3D environment with a point transmitter and a passive spherical receiver of radius $\rrx$ that is centered at a distance $\dist$ from the transmitter. When the transmitter releases a molecule at time $t=0$, the probability that the molecule is inside the receiver for the $\obsInd$th observation is accurately approximated as \cite[Eq.~(3.5)]{Crank1979}
	\begin{equation}
		\probSingleObs{\obsInd} =  \frac{\Vrx}{(4\pi\diffusion \obsInd\dt)^\frac{3}{2}}\EXP{-\frac{\dist^2}{4\diffusion \obsInd\dt}},
		\label{eqn_cir}
	\end{equation}
	where $\diffusion$ is the constant diffusion coefficient and $\Vrx$ is the receiver volume.
	
	For evaluation, we consider the system parameters that are summarized in Table~\ref{table_param}. The transmitter modulates a sequence of $\dataLength=20$ binary symbols that are generated randomly and with equal probability. $\molReleased=2\times10^4$ molecules are released for each bit-1. The corresponding average channel response from a single bit-1 is plotted in Fig.~\ref{fig_cir} on page~\pageref{fig_cir}. The simulation is repeated $10^3$ times, so there are a total of $\molReleased=2\times10^4$ transmitted bits. The expected average error probability $\pErrorAvg$ is calculated from the bit sequences that were simulated; for each sequence, we calculate the expected error probability of every bit, and then we take the average over all bits and all sequences.
	
	\begin{table}[!tb]
		\centering
		\caption{Simulation system parameters.}
		
		{\renewcommand{\arraystretch}{1.4}
			\begin{tabular}{l|c|c||c}
				\hline
				\bfseries Parameter & \bfseries Symbol & \bfseries Units& \bfseries Value \\ \hline \hline
				RX Radius & $\rrx$ &	$\mu\metre$
				& 0.5  \\ \hline
				Molecules Released & $\molReleased{}$ & $\mol$
				& $2\times10^4$ \\ \hline
				Sequence Length & $\dataLength{}$ & -
				& 20 \\ \hline
				Distance to RX & $\dist{}$ & $\mu\metre$
				& 5 \\ \hline
				Diffusion Coeff.
				& $\diffusion{}$ & $\metre^2/\second$
				& $10^{-10}$ \\ \hline
				Sampling Period & $\dt$ & $\metre\second$
				& $\{8,40\}$ \\ \hline
				Symbol Period & $\obsPerBit\dt$ & $\metre\second$
				& 200 \\ \hline
				\# of Realizations & - & -
				& $10^3$ \\ \hline
			\end{tabular}
		}
		\label{table_param}
	\end{table}
	
	Overall, the simulation and analytical curves in this section agree very well, and slight deviations (particularly at lower error probabilties) are primarily because we calculated the \emph{average} error probability of every sequence but only simulated each sequence \emph{once}. We also note that the bit error probabilities observed throughout this section are relatively high (i.e., generally above $10^{-3}$). This was deliberately imposed so that we could accurately observe meaningful bit error probabilities from $2\times10^4$ bits for all detectors considered. Specifically, the symbol period of $200\,\metre\second$ is not significantly larger than the expected peak time; from Fig.~\ref{fig_cir}, the average number of molecules expected after $200\,\metre\second$ is still about one third the number expected at the expected peak time of $\dist^2/(6\diffusion)\approx40\,\metre\second$. Furthermore, the individual observations are relatively small; the peak of the expected channel response is only 6 molecules. Methods to achieve lower bit error probabilities include using a larger symbol period and (as we will see in Section~\ref{sec_results_samples}) sampling more often.
	
	\subsection{Error Versus Threshold}
	
\begin{figure}[!tb]
		\centering
		\includegraphics[width=\linewidth]{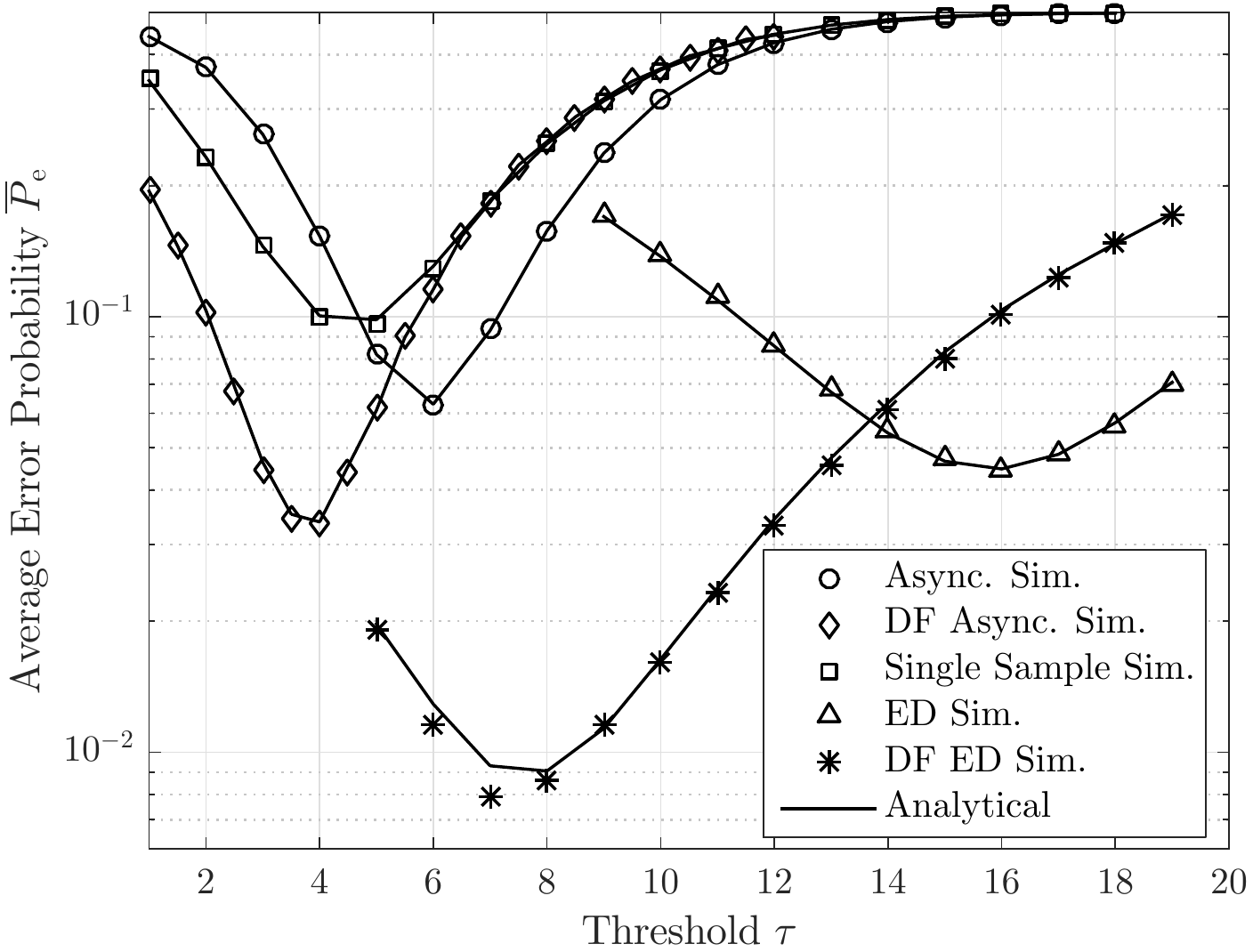}
		\caption{Average probability of error for different threshold $\threshold$. The sampling period is $\dt = 40\,$ms. The asynchronous (Async.) detector is compared with the single sample detector and the energy detector (ED). Asynchronous and energy detectors are also considered with decision feedback (DF).}
		\label{fig_error_vs_thresh_40ms}
	\end{figure}

	First, we consider the case where the receiver sampling is synchronized with the symbol intervals of the transmitter, i.e., $\obsOffset = 0$. We vary the constant (baseline) decision threshold $\threshold$ and measure the corresponding average bit error probability $\pErrorAvg$ for the detectors considered. We compare the detectors in Fig.~\ref{fig_error_vs_thresh_40ms}. As expected, every curve shows an increasing error probability with sufficiently high and sufficiently low $\threshold$, such that there exists an optimal threshold that balances the correct detection of bit-1 versus the correction detection of bit-0. Analytical derivations for approximating the optimal threshold of the existing detectors were presented in \cite{Tepekule2015a,Ahmadzadeh2015a}.
	
	In Fig.~\ref{fig_error_vs_thresh_40ms}, we consider a sampling period of $\dt = 40\,\metre\second$ (i.e., there are $\obsPerBit=5$ samples per bit). The single sample detector uses the 1st sample in every interval for detection. The best error probability observed for the single sample detector, which is above $0.09$, is higher than that for the asynchronous detector, which is less than $0.07$. This improvement comes at the cost of taking the maximum of all 5 samples, but the key advantage is that the asynchronous detector did not need know \emph{when} the maximum sample was expected. Additional performance improvement is observed by adding decision feedback (DF) to the asynchronous detector, which decreases the minimum error probability to below $0.05$. This adaptive asynchronous detector is actually better than the (non-adaptive) energy detector. However, adding decision feedback to the energy detector leads to a significant improvement, such that its minimum error probability drops to about $0.008$. Overall, in the synchronized communication case, the performance of the asynchronous detector ranks between that of the single sample detector and that of the energy detector.
	
	\subsection{Error Versus Receiver Clock Offset}
	\label{sec_results_offset}
	
\begin{figure}[!tb]
		\centering
		\includegraphics[width=\linewidth]{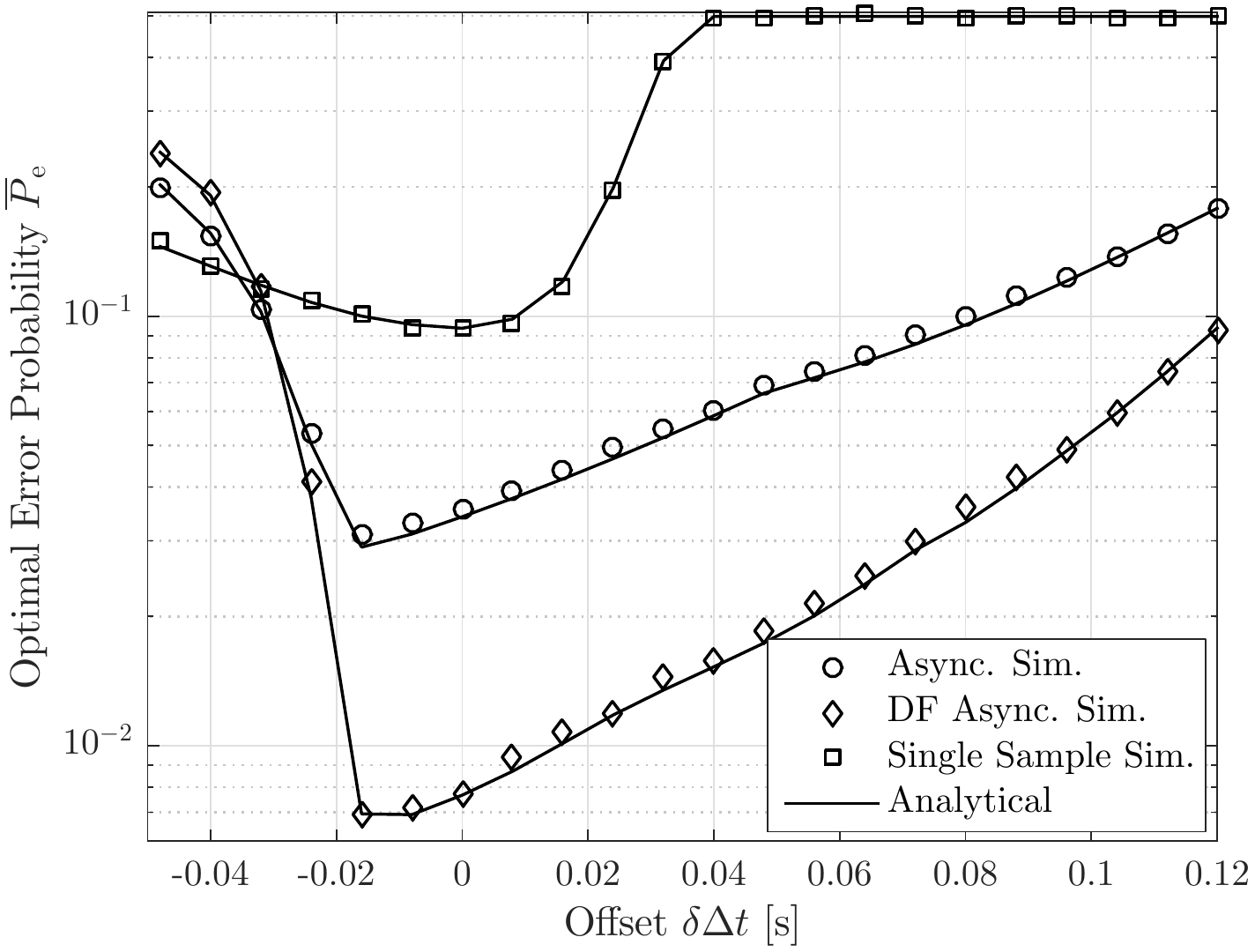}
		\caption{Average probability of error at optimal threshold for different sampling offsets $\obsOffset$. The sampling period is $\dt = 8\,$ms. The asynchronous detectors are less sensitive to positive $\obsOffset$ than the single sample detector.}
		\label{fig_error_vs_offset_8ms}
	\end{figure}

\begin{figure}[!tb]
		\centering
		\includegraphics[width=\linewidth]{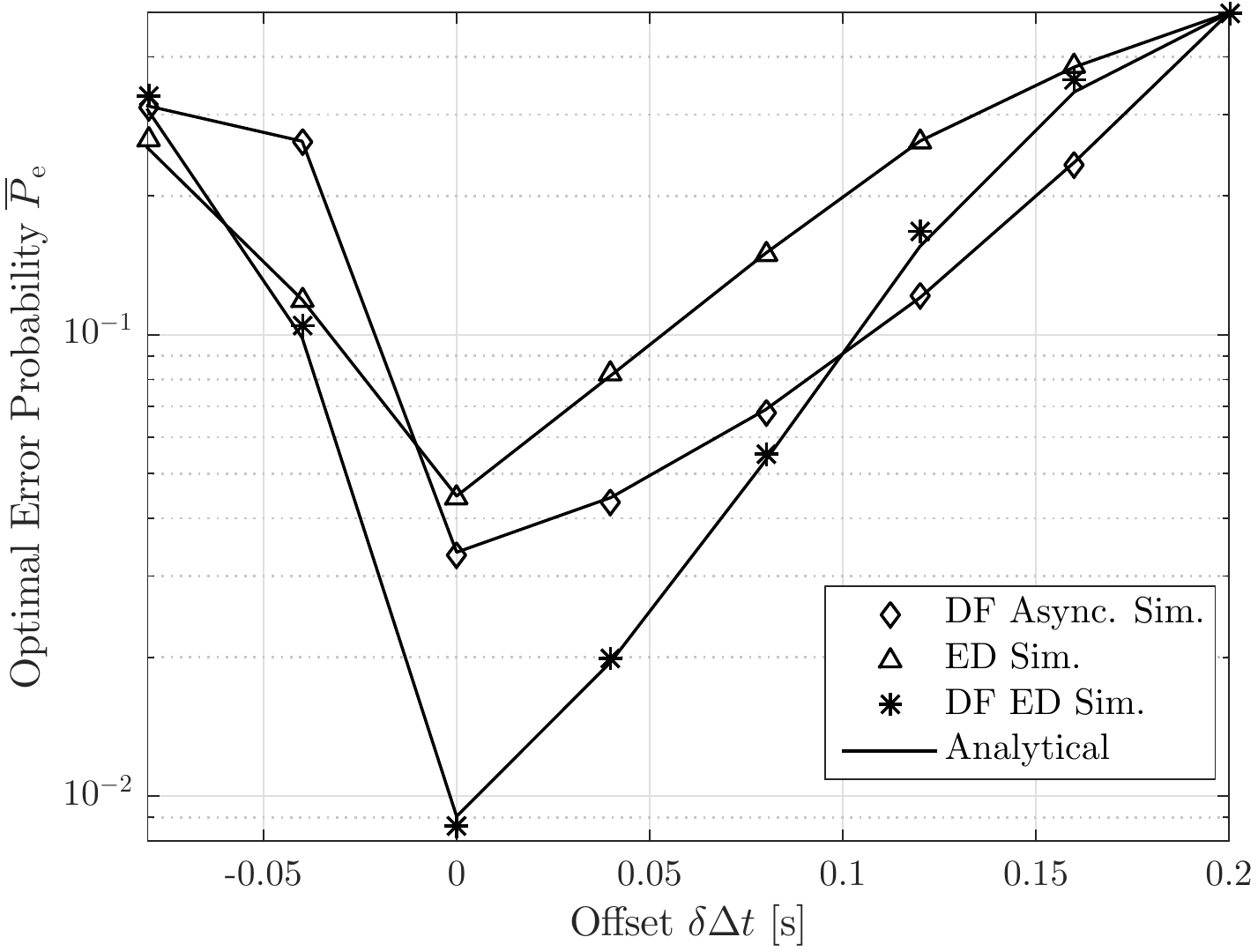}
		\caption{Average probability of error at the optimal threshold for different sampling offsets $\obsOffset$. The sampling period is $\dt = 40\,$ms. The asynchronous detector with decision feedback is less sensitive to positive $\obsOffset$ than the energy detectors.}
		\label{fig_error_vs_offset_40ms}
	\end{figure}

	Second, we vary the timing offset $\obsOffset$ and measure the corresponding minimum bit error probability. For a given offset, we use the analytical expression to numerically find the threshold $\threshold$ that gives the minimum expected error probability, and then we measure the average error probability from the simulations using that threshold. For simplicity, we assume that the signal $\obsSignal{\obsIndRX}$ has value zero beyond the intended sampling interval, i.e., $\obsSignal{\obsIndRX} = 0, \obsIndRX \notin \{0,1,\ldots,\obsPerBit\dataLength-1\}$. For ease of comparison, and to demonstrate lower error probabilities for the asynchronous detectors, the results are presented in Figs.~\ref{fig_error_vs_offset_8ms} and \ref{fig_error_vs_offset_40ms} for sampling intervals of $\dt=8\,\metre\second$ and $\dt=40\,\metre\second$, respectively. Furthermore, we only consider integer values of $\obsOffset$.
	
	In Fig.~\ref{fig_error_vs_offset_8ms}, we consider sample offset $\obsOffset \in [-6,15]$, i.e., the offset between the receiver's and transmitter's clocks is between $-48\,\metre\second$ and $120\,\metre\second$. The performance of the single sample detector is highly sensitive to positive $\obsOffset$, because the expected peak time is relatively early in the symbol interval. When the offset is $\ge 40\,\metre\second$, this detector is no longer sampling in the intended symbol interval and its performance is very poor. The asynchronous detectors (both with and without decision feedback) are much more resilient to positive offset $\obsOffset$ and demonstrate a slow performance decay as $\obsOffset$ increases. Interestingly, these detectors actually \emph{improve} with a small negative $\obsOffset$, because in the first few observations of a symbol interval when $\obsOffset=0$ we are more likely to see molecules due to the previous symbol than due to the current symbol (see Fig.~\ref{fig_cir}). However, performance rapidly degrades as $\obsOffset$ decreases further and the receiver's largest observation is more likely to include molecules that were released for the following symbol.
	
	In Fig.~\ref{fig_error_vs_offset_40ms}, we consider sample offset $\obsOffset \in [-2,5]$, i.e., the offset between the receiver's and transmitter's clocks is between $-40\,\metre\second$ and $200\,\metre\second$. For negative $\obsOffset$, the performance of all detectors degrades significantly. However, the asynchronous detector with decision feedback is the \emph{least sensitive} to positive $\obsOffset$, and it even performs \emph{better} than the energy detector with decision feedback when the offset is greater than $100\,\metre\second$ (albeit with an error probability above $0.1$). This is because the energy detector is including a lot of energy from the ``tail'' of the previous symbol in its bit decision, whereas the asynchronous detector is only comparing the value of the largest (adaptive) sample and this maximum is more likely to come from the intended symbol interval (e.g., consider the left half vs the right half of Fig.~\ref{fig_cir}).
	
	\subsection{Error Versus Sampling Period}
	\label{sec_results_samples}
	
\begin{figure}[!tb]
		\centering
		\includegraphics[width=\linewidth]{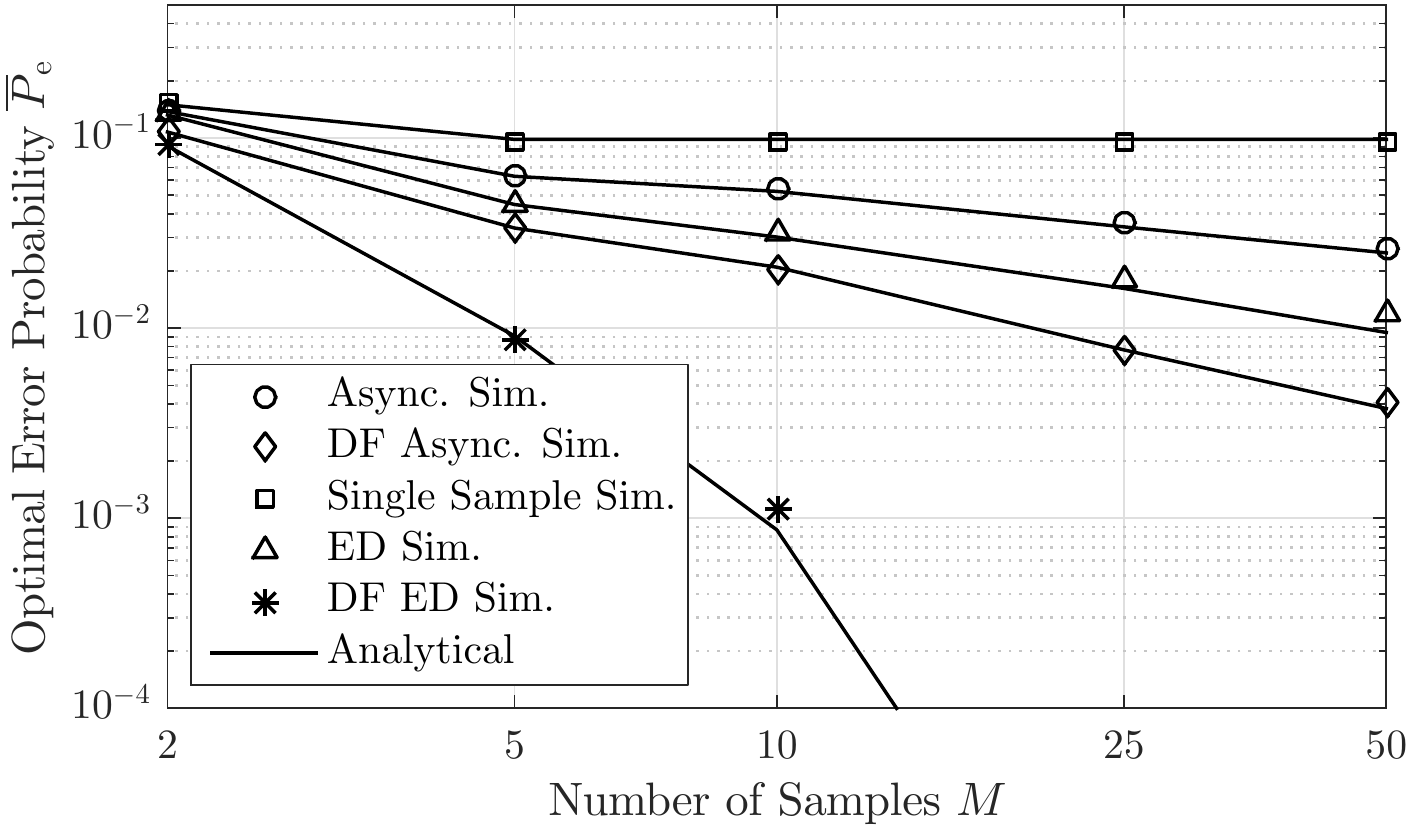}
		\caption{Average probability of error at the optimal threshold for different numbers of samples per interval $\obsPerBit$. The sampling period $\dt$ is $200\,\metre\second/\obsPerBit$. All detectors except the single sample detector improve with increasing $\obsPerBit$. Error probabilities for the energy detector with decision feedback are not shown for $\obsPerBit > 10$ because only $2\times10^4$ bits were simulated.}
		\label{fig_error_vs_samples}
	\end{figure}

	Finally, we explicitly study the impact of the number of samples $\obsPerBit$ on the error probability, where the minimum probability for a given number of samples is determined using the same method as that described for a given offset in Section~\ref{sec_results_offset}. Here, we set the offset to $\obsOffset = 0$ and consider $\obsPerBit=\{2,5,10,25,50\}$. The results are shown in Fig.~\ref{fig_error_vs_samples}. For $\obsPerBit>5$, the performance of the single sample detector does not improve since the sampling time stays the same. All other detectors, including the asynchronous detectors, improve with increasing $\obsPerBit$ over the range considered. The energy detector with decision feedback actually improves by many orders of magnitude, although in practice detector performance would eventually be limited by sample dependence.
	
	\section{Conclusions}
	\label{sec_concl}
	
	In this paper, we proposed and analyzed the asynchronous peak detector, both with and without decision feedback. Both variants demonstrate resilience to timing offsets between the transmitter and receiver, and can outperform existing detectors. Future work in this area includes: 1) modeling sample dependence in the distribution of the largest observation; and 2) optimizing the symbol sampling window.

	\bibliographystyle{IEEEtran}
	
	\bibliography{../references/library_fixed}

\begin{thebibliography}{10}
\providecommand{\url}[1]{#1}
\csname url@samestyle\endcsname
\providecommand{\newblock}{\relax}
\providecommand{\bibinfo}[2]{#2}
\providecommand{\BIBentrySTDinterwordspacing}{\spaceskip=0pt\relax}
\providecommand{\BIBentryALTinterwordstretchfactor}{4}
\providecommand{\BIBentryALTinterwordspacing}{\spaceskip=\fontdimen2\font plus
\BIBentryALTinterwordstretchfactor\fontdimen3\font minus
  \fontdimen4\font\relax}
\providecommand{\BIBforeignlanguage}[2]{{%
\expandafter\ifx\csname l@#1\endcsname\relax
\typeout{** WARNING: IEEEtran.bst: No hyphenation pattern has been}%
\typeout{** loaded for the language `#1'. Using the pattern for}%
\typeout{** the default language instead.}%
\else
\language=\csname l@#1\endcsname
\fi
#2}}
\providecommand{\BIBdecl}{\relax}
\BIBdecl

\bibitem{Nakano2013c}
T.~Nakano, A.~W. Eckford, and T.~Haraguchi, \emph{Molecular
  Communication}.\hskip 1em plus 0.5em minus 0.4em\relax Cambridge University
  Press, 2013.

\bibitem{Abadal2011a}
S.~Abadal and I.~F. Akyildiz, ``Bio-inspired synchronization for
  nanocommunication networks,'' in \emph{Proc. IEEE GLOBECOM}, 2011, pp.
  5375--5379.

\bibitem{Moore2013}
M.~J. Moore and T.~Nakano, ``Oscillation and synchronization of molecular
  machines by the diffusion of inhibitory molecules,'' \emph{IEEE Trans.
  Nanotechnol.}, vol.~12, no.~4, pp. 601--608, Jul. 2013.

\bibitem{Shahmohammadian2013}
H.~Shahmohammadian, G.~G. Messier, and S.~Magierowski, ``Blind synchronization
  in diffusion-based molecular communication channels,'' \emph{IEEE Commun.
  Lett.}, vol.~17, no.~11, pp. 2156--2159, Nov. 2013.

\bibitem{Lin2016}
L.~Lin, C.~Yang, M.~Ma, S.~Ma, and H.~Yan, ``A clock synchronization method for
  molecular nanomachines in bionanosensor networks,'' \emph{IEEE Sens. J.},
  vol.~16, no.~19, pp. 7194--7203, 2016.

\bibitem{Moore2012}
M.~J. Moore, T.~Nakano, A.~Enomoto, and T.~Suda, ``Measuring distance from
  single spike feedback signals in molecular communication,'' \emph{IEEE Trans.
  Signal Process.}, vol.~60, no.~7, pp. 3576--3587, Jul. 2012.

\bibitem{Noel2014c}
A.~Noel, K.~C. Cheung, and R.~Schober, ``Joint channel parameter estimation via
  diffusive molecular communication,'' \emph{IEEE Trans. Mol. Biol. Multi-Scale
  Commun.}, vol.~1, no.~1, pp. 4--17, Mar. 2015.

\bibitem{Rose2015}
C.~Rose and I.~S. Mian, ``A fundamental framework for molecular communication
  channels: Timing payload,'' in \emph{Proc. IEEE ICC}, London, United kingdom,
  Jun. 2015, pp. 1043--1048.

\bibitem{Srinivas2012}
K.~V. Srinivas, A.~W. Eckford, and R.~S. Adve, ``Molecular communication in
  fluid media: The additive inverse gaussian noise channel,'' \emph{IEEE Trans.
  Inf. Theory}, vol.~58, no.~7, pp. 4678--4692, Jul. 2012.

\bibitem{Lee2015a}
J.~Lee, M.~Kim, and D.-H. Cho, ``Asynchronous detection algorithm for
  diffusion-based molecular communication in timing modulation channel,''
  \emph{IEEE Commun. Lett.}, vol.~19, no.~12, pp. 2114--2117, Dec. 2015.

\bibitem{Lin2015b}
Y.-K. Lin, W.-A. Lin, C.-H. Lee, and P.-C. Yeh, ``Asynchronous threshold-based
  detection for quantity-type-modulated molecular communication systems,''
  \emph{IEEE Trans. Mol. Biol. Multi-Scale Commun.}, vol.~1, no.~1, pp. 37--49,
  Mar. 2015.

\bibitem{Akyildiz2008}
I.~F. Akyildiz, F.~Brunetti, and C.~Bl{\'{a}}zquez, ``Nanonetworks: A new
  communication paradigm,'' \emph{Comput. Networks}, vol.~52, no.~12, pp.
  2260--2279, Aug. 2008.

\bibitem{Li2016a}
B.~Li, M.~Sun, S.~Wang, W.~Guo, and C.~Zhao, ``Local convexity inspired
  low-complexity noncoherent signal detector for nanoscale molecular
  communications,'' \emph{IEEE Trans. Commun.}, vol.~64, no.~5, pp. 2079--2091,
  May 2016.

\bibitem{Damrath2016}
M.~Damrath and P.~A. Hoeher, ``Low-complexity adaptive threshold detection for
  molecular communication,'' \emph{IEEE Trans. Nanobioscience}, vol.~15, no.~3,
  pp. 200--208, Apr. 2016.

\bibitem{Hong2013}
Y.~Hong, ``On computing the distribution function for the {Poisson} binomial
  distribution,'' \emph{Comput. Stat. Data Anal.}, vol.~59, pp. 41--51, Mar.
  2013.

\bibitem{Ross2009}
S.~M. Ross, \emph{Introduction to Probability and Statistics for Engineers and
  Scientists}, 4th~ed.\hskip 1em plus 0.5em minus 0.4em\relax Academic Press,
  2009.

\bibitem{Abramowitz1964}
M.~Abramowitz and I.~A. Stegun, \emph{Handbook of Mathematical Functions with
  Formulas, Graphs, and Mathematical Tables}, 1st~ed.\hskip 1em plus 0.5em
  minus 0.4em\relax United States Department of Commerce, National Bureau of
  Standards, 1964.

\bibitem{Noel2014d}
A.~Noel, K.~C. Cheung, and R.~Schober, ``Optimal receiver design for diffusive
  molecular communication with flow and additive noise,'' \emph{IEEE Trans.
  Nanobiosci.}, vol.~13, no.~3, pp. 350--362, Sep. 2014.

\bibitem{Noel2014e}
------, ``Overcoming noise and multiuser interference in diffusive molecular
  communication,'' in \emph{Proc. ACM NANOCOM}, 2014, pp. 1--9.

\bibitem{Noel2016}
\BIBentryALTinterwordspacing
A.~Noel, ``{AcCoRD} ({A}ctor-based {C}ommunication via
  {R}eaction-{D}iffusion).'' [Online]. Available:
  \url{https://github.com/adamjgnoel/AcCoRD/}
\BIBentrySTDinterwordspacing

\bibitem{Crank1979}
J.~Crank, \emph{The Mathematics of Diffusion}, 2nd~ed.\hskip 1em plus 0.5em
  minus 0.4em\relax Oxford University Press, 1979.

\bibitem{Tepekule2015a}
B.~Tepekule, A.~E. Pusane, H.~B. Yilmaz, C.-B. Chae, and T.~Tugcu, ``{ISI}
  mitigation techniques in molecular communication,'' \emph{IEEE Trans. Mol.
  Biol. Multi-Scale Commun.}, vol.~1, no.~2, pp. 202--216, Jun. 2015.

\bibitem{Ahmadzadeh2015a}
A.~Ahmadzadeh, A.~Noel, and R.~Schober, ``Analysis and design of multi-hop
  diffusion-based molecular communication networks,'' \emph{IEEE Trans. Mol.
  Biol. Multi-Scale Commun.}, vol.~1, no.~2, pp. 144--157, Jun. 2015.

\end{thebibliography}
	
\end{document}